\documentclass[10pt,letterpaper,twocolumn]{article} %% two column, final layout

\usepackage{ol2}
\usepackage[draft]{hyperref}
\usepackage{amsmath}

\begin{document}

\twocolumn[ %% activate for two-column option

\title{Dual-wavelength laser source for onboard atom interferometry}

%% For REVTeX it is possible to automate superscript and e-mail callouts with the superscriptaddress option; see REVTeX4 documentation.

\author{V. M\'enoret$^{1,*}$, R. Geiger$^{1,4}$, G. Stern$^{1,4}$, N. Zahzam$^{3}$, B. Battelier$^{1,5}$, A. Bresson$^{3}$, A. Landragin$^{2}$, P. Bouyer$^{1,5}$}

\address{
$^1$Laboratoire Charles Fabry, IOGS, CNRS and Universit\'e Paris Sud 11, 2 avenue Fresnel, 91127 Palaiseau, France\\
$^2$LNE-SYRTE, Observatoire de Paris, CNRS and UPMC, 61 avenue de l'Observatoire, 75014 Paris, France\\
$^3$ONERA - The French Aerospace Lab, F-91761 Palaiseau, France\\
$^4$Centre National d'Etudes Spatiales, 18 avenue Edouard Belin, 31401 Toulouse, France\\
%$^5$Laboratoire Photonique Num\'erique et Nanosciences, Universit\'e Bordeaux 1, IOGS and CNRS, 351 Cours de la Lib\'eration, 33405 Talence, France\\
$^5$LP2N, Universit\'e Bordeaux 1, IOGS and CNRS, 351 Cours de la Lib\'eration, 33405 Talence, France\\
$^*$Corresponding author: vincent.menoret@institutoptique.fr
}

\begin{abstract}
We present a compact and stable dual-wavelength laser source for onboard atom interferometry with two different atomic species. It is based on frequency-doubled telecom lasers locked on a femtosecond optical frequency comb. We take advantage of the maturity of fiber telecom technology to reduce the number of free-space optical components which are intrinsically less stable, and to make the setup immune to vibrations and thermal fluctuations. The source provides the frequency agility and phase stability required for atom interferometry and can easily be adapted to other cold atom experiments. We have shown its robustness by achieving the first dual-species K-Rb magneto optical trap in microgravity during parabolic flights.\end{abstract}


 ] %% activate for two-column option

\noindent Atom interferometers have demonstrated excellent performances for precision acceleration and rotation measurements \cite{gyro}. Many foreseen applications of these instruments such as tests of fundamental physics \cite{dimopoulos, quantus} or terrain gravimetry \cite{mcguirk} require the setup to be transportable and able to operate in harsh environments. A lot of efforts have been made in the past few years to develop such transportable systems \cite{syrte,Peters,sorrentino}, and in particular to make the laser setup as compact and immune to perturbations as possible \cite{remi}.

Atom Interferometers usually operate with alkali atoms by driving transitions in the near-infrared spectrum (852 nm for Cs, 780 nm for Rb and 767 nm for K). In these experiments one needs stable lasers both to cool the atoms and to interrogate them by using stimulated Raman transitions \cite{kasevich91}. These Raman transitions require two phase-locked lasers red-detuned by approximately 1 GHz with respect to the cooling wavelength, and with a frequency separation corresponding to the ground state hyperfine splitting of the atom. The performance of the phase-lock is critical since it has a direct impact on the interferometric measurement.

While most atom interferometers use laser systems based on extended-cavity diode lasers (ECDLs) \cite{syrtelaser}, a good alternative is to use frequency-doubled telecom lasers operating around 1.5 $\mu$m \cite{carraz, MOT_K}. This technique takes advantage of the maturity of fiber components in the telecom C-band to reduce the amount of free-space optics and to make the setup more compact and less sensitive to mis-alignments. The second Raman frequency is generated by modulating the reference laser with an electro-optic or acousto-optic modulator, thus avoiding phase-locking of two diodes.

We present a new laser source for onboard atom interferometry with two species. It is based on two telecom Distributed Feedback (DFB) lasers at 1560 and 1534 nm that are combined, amplified and frequency-doubled to 780 and 767 nm in order to manipulate Rb and K atoms simultaneously. Both DFBs are frequency-locked on a fiber-based optical frequency comb (FOFC) \cite{femto}.

\begin{figure}[htb]
\centerline{\includegraphics[width=8.3cm]{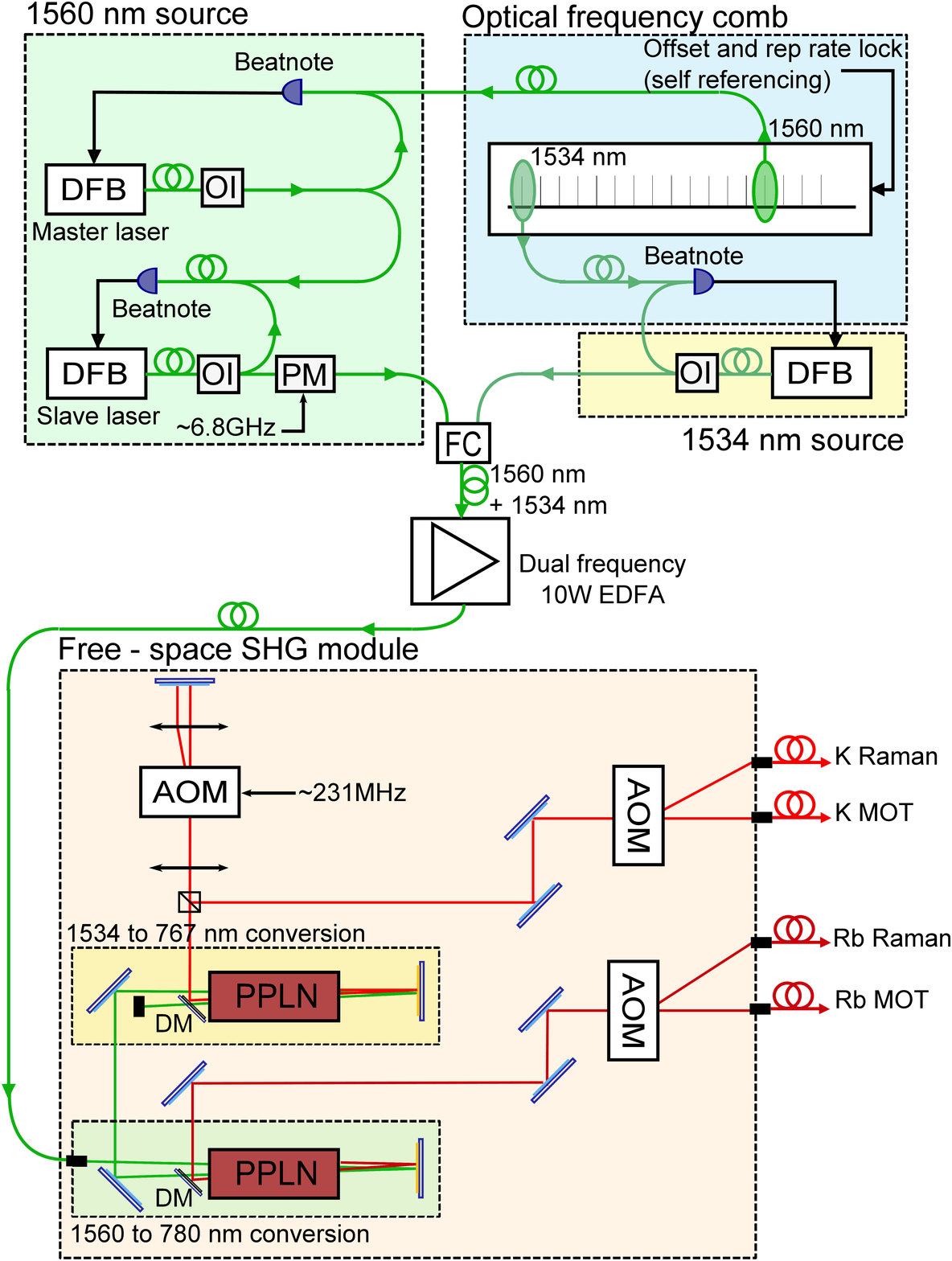}}
\caption{Diagram of the complete laser system. DFB : Distributed Feedback laser diode. OI : Optical Isolator. PM : Phase Modulator. FC : Fiber combiner. SHG : Second Harmonic Generation. DM : Dichroic Mirror. PPLN : Periodically Poled Lithium Niobate crystal. AOM : Acousto-Optic Modulator.}
\label{setup}
\end{figure}

We have demonstrated the robustness of this source during parabolic flight campaigns in which we obtained for the first time a double-species magneto-optical trap (MOT) in microgravity aboard the Novespace zero-g aircraft \cite{Novespace}. The system was operated despite temperature fluctuations from 17 to 30$^{\circ} \mathrm{C}$, high vibration levels up to 0.5 m.s$^{-2}$ rms and a vertical acceleration alternating between 0, 10 and 20 m.s$^{-2}$. Apart from a free-space module all the components are fibered and fit in standard 19'' rack structures. The source meets the usual requirements for atom interferometry (Fig. \ref{setup}) : separate optical outputs for MOT and Raman lasers with powers of the order of 200 mW, frequency agility over hundreds of MHz, AOM-driven microsecond pulses  and high phase stability for the Raman lasers. In the following, we first detail the optical characteristics of the source, then we focus on the phase stability of the lasers and describe the dual-species MOT demonstrated in microgravity.

We stabilize the Rb and K DFB lasers on a common frequency reference by the means of a FOFC. This technique has been preferred over conventional saturated absorption spectroscopy methods because it offers a precise measurement of the absolute optical frequencies, along with common mode frequency noise rejection. The comb has been specifically designed by MenloSystems to be transportable and resistant to vibrations. It is self-referenced on a 10 MHz ultra-stable quartz oscillator which is the RF frequency reference for the whole atom interferometry experiment \cite{Stern}. Built-in fibered beat detection units allow the detection of a beat signal between the comb and each of the DFB lasers, and the stabilization of these lasers on the comb.

The 1560 nm source is inspired from that described in \cite{Stern}. A master DFB laser is locked on the FOFC at a fixed frequency by retroacting on the diode current. A slave DFB is then locked on the master laser with an adjustable setpoint, allowing frequency detuning at 1560 nm over 500 MHz in approximately 3 ms. This agility is needed in order to detune the Raman laser for the interferometer sequence. The slave laser is then used for both cooling and interrogating $^{87}$Rb. The 1534 nm source is made of a single DFB locked on the FOFC\cite{raman_K}.

The two lasers at 1560 and 1534 nm are combined in the same fiber and amplified to approximately 5W each in a dual-frequency Erbium-Doped Fiber Amplifier manufactured by Keopsys, designed so as to have the same gain at the two wavelengths. By finely adjusting the value of the two input powers we can set the power ratio on the output light to the desired value.

The amplified light is coupled out and frequency-doubled in free-space. We use two double-pass crystals in series to frequency-double the two lasers at 1560 and 1534 nm respectively (see Fig. 1). The light first goes through a 3 cm long PPLN crystal from HCPhotonics phase-matched for the Second Harmonic Generation of 780 nm light starting from 1560 nm. We extract the 780 nm component on a dichroic mirror and send the remaining 1.5 $\mu$m light through a second PPLN crystal phase-matched for the conversion of 1534 to 767 nm light.The double-pass configuration enables high efficiencies of approximately 4\%.W$^{-1}$ with 5W pump power.

We generate the repumping and second Raman frequencies by modulating the lasers at frequencies close to the hyperfine splittings of $^{87}$Rb and $^{40}$K (6.8 GHz and 462 MHz respectively). On the Rubidium laser we use a fibered phase modulator operating at 1560 nm and driven by a 6.8 GHz signal from our frequency reference. The parasitic sidebands generated by the phase modulator are not critical on the Rubidium setup because they are far away from the atomic transition and have little effect. On Potassium the hyperfine splitting is smaller and phase modulation would lead to spurious sidebands close to resonance, causing enhanced spontaneous emission and parasitic interferometers. We therefore use a free-space acousto-optic modulator (AOM) in a double pass configuration at 767 nm. We superimpose the fundamental order and the one diffracted twice in the AOM so as to get two frequencies separated by 462 MHz.

Each of the two frequency-doubled lasers is sent through an AOM in order to be split in two paths, one for the cooling beam and one for the Raman beam. All 4 paths (2 Rb and 2 K) go through fast mechanical shutters (Uniblitz LS2, 300 $\mu$s rise time) and are coupled in polarization-maintaining fibers. The fiber-coupling efficiency is of the order of 70\%, and we get around 200 mW optical power at each of the fiber outputs. In addition, a few mW of light of each wavelength is used to generate blowaway beams.

Since the free-space bench is a critical part of the setup we have taken special care to make it as stable as possible. Our experiment is operated in a plane carrying out parabolic flights so the setup has to withstand high levels of vibration, as well as local gravity changes. The optics are screwed on a custom made AW2618 aeronautic grade aluminium breadboard which is more stable than standard AW2017. The breadboard is 43x73 cm$^2$ and 4 cm thick, with a hollow structure to make it lighter.

\begin{figure}[tb]
\centerline{\includegraphics[width=8.3cm]{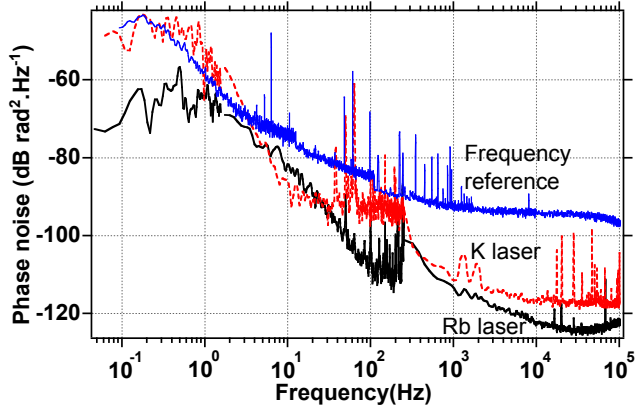}}
\caption{Phase noise due to modulation and propagation for the two lasers compared to that of our 6.8 GHz frequency reference. For the Rb laser, phase noise is dominated by that of the reference. On the K laser excess noise appears at low frequency because of physical separation between the two frequency components.}
\label{phase_noise}
\end{figure}

We have measured the phase noise due to propagation on the 780 (resp. 767) nm Raman lasers by detecting a beatnote at 6.8 GHz (resp. 462 MHz) on a fast photodiode, and mixing it with the same signal that was used for the sideband generation. On the 780 nm Rb laser the phase noise resulting from the phase modulator and propagation in the system is more than one order of magnitude lower than that of the frequency reference itself, meaning that our solution is a good alternative to phase-locked diode lasers (Fig.\ \ref{phase_noise}). The 767 nm source shows similar performance in quiet conditions, but is more sensitive to low frequency-phase noise due to vibrations and air flows because of the physical path separation induced by the double-pass AOM. However, even under strong mechanical constraints the RMS phase noise remains below 200 mrad, and below 30 mrad in quiet conditions. Further stabilization can be added by retroacting on the phase of the RF signal driving the AOM.

Finally, the laser source has been validated during parabolic flight campaigns organized by CNES and ESA, by achieving the first dual-species MOT in microgravity. We capture around $10^9$ Rb atoms and $5.10^7$ K atoms in a static MOT, loaded from a vapor. The FOFC has remained fully locked and stable throughout the flights. The output optical powers were also stable, with no noticeable long-term degradation or influence of vibrations ($< 5\%$). The temperature of Rubidium atoms after sub-Doppler cooling in optical molasses is of the order of 8 $\mu$K, allowing high resolution acceleration measurements in the plane \cite{remi}. We have measured the temperature of K atoms by a release and recapture method, and estimate it around 200 $\mu$K, which is comparable to what has been obtained in \cite{MOT_K_italie}. Further cooling to 25 $\mu$K in optical molasses has been recently achieved \cite{fattori,MOT_K_Inde} and will be implemented soon on our experiment.

Improvements to this laser source could come from the reduction of free-space optics. For instance, the use of pigtailed PPLN waveguides instead of bulk crystals could lead to a smaller setup, along with a reduction of the needed pump power. Studies in this direction are currently under way and promising results have been obtained by Nishikawa and coworkers \cite{PPLN_WG}. Furthermore, fast and polarization maintaining switches with high extinction rations operating at 780 or 767 nm are difficult to find. The maturation of these technologies seems necessary before a fully fibered laser source for atom interferometry can be built. 

The laser source we have presented is a compact and versatile setup, designed to test the weak Equivalence Principle in the zero-g plane \cite{gael}. A similar design could easily be adapted to other dual-species cold atom experiments, the use of the frequency comb allowing for easy absolute frequency determination and stabilization. The high stability and reliability of this type of fiber-based architecture also make it a good alternative to ECDL-based systems for laboratory experiments.

We thank L. Mondin and T. L\'ev\`eque for their implication in the project. We acknowledge funding supports from CNES, DGA, CNRS, ESA and IFRAF.

%\pagebreak

%\bibliographystyle{ol}
%\bibliography{biblio}
%\bibliographystyle{osajnl}
%\bibliography{biblio}
%\end{document}

%%\bibliographystyle{ol}
%%\bibliography{biblio}

\pagebreak
\section*{Informational Fourth Page}
%%In this section, please provide full versions of citations to assist reviewers and editors (OL publishes a short form of citations) or any other information that would aid the peer-review process.

%%\bibliographystyle{osajnl}
%%\bibliography{biblio}

\begin{thebibliography}{10}
\bibitem{raman_K} 
  A master/slave architecture similar to that of Rubidium will be implemented for Raman detuning.
\end{thebibliography}
\begin{thebibliography}{10}
\bibitem{raman_K}
A master/slave architecture similar to that of rubidium will be
  implemented for raman detuning.
\end{thebibliography}

\end{document}